\documentclass[review]{article}
\usepackage{amsmath,amssymb,amsfonts,amsthm,fullpage,setspace}

\newtheorem{theorem}{Theorem}
\newtheorem{lemma}{Lemma}
\newtheorem{corollary}{Corollary}

\title{Complexity of Computing the Projection of Polytopes}
\author{Hans Raj Tiwary\\
hansraj@cs.uni-sb.de
}

\date{}
\begin{document}
\maketitle

\begin{abstract}
We study the complexity of computing the projection of an arbitrary $d$-polytope along $k$ orthogonal vectors for various input and output forms. We show that if $d$ and $k$ are part of the input (\emph{i.e.} not a constant) and we are interested in output-sensitive algorithms, then in most forms the problem is equivalent to enumerating vertices of polytopes, except in two where it is NP-hard. In two other forms the problem is trivial. We also review the complexity of computing projections when the projection directions are in some sense non-degenerate. For full-dimensional polytopes containing origin in the interior, projection is an operation dual to intersecting the polytope with a suitable linear subspace and so the results in this paper can be dualized by interchanging vertices with facets and projection with intersection. To compare the complexity of projection and vertex enumeration, we define new complexity classes based on the complexity of Vertex Enumeration.
\end{abstract}

\section{Introduction}
A polytope in $\mathbb{R}^d$ is a closed convex body that can be represented as either the convex hull of a finite number of points or as the intersection of a finite number of halfspaces. We will call the former $\mathcal{V}$-representation and the latter $\mathcal{H}$-representation. Accordingly a polytope will be called a $\mathcal{V}$-polytope, an $\mathcal{H}$-polytope or an $\mathcal{HV}$-polytope depending on whether the polytope is given by $\mathcal{V}$, $\mathcal{H}$ or both representations. For any polytope each of these representations is unique if no redundancies are allowed and any of these representations completely determines the others. We refer the reader to \cite{grunbaum, ziegler_book} for a thorough treatment of the subject.

For enumeration problems, it is often the case that the size of the output is not bounded by a polynomial in the input size, and hence any algorithm computing the desired output can not have polynomial running time. Therefore it is useful to talk about output-sensitive algorithms whose running time is measured in terms of both the size of the input and the output. In this paper we consider only output-sensitive algorithms and when we talk about polynomiality of the algorithm we implicitly assume output-sensitivity.

The Vertex Enumeration problem (VE) asks one to enumerate the vertices of a polytope given by its facets. Despite having been studied for a long time by a number of researchers, the complexity status of VE, for general dimension and for polytopes that are neither simple nor simplicial, is unknown. It is neither known to be NP-hard nor there exists any polynomial algorithm for it. The dual problem of computing $\mathcal{H}$-representation from $\mathcal{V}$-representation is known as Convex Hull problem (CH). These two problems are equivalent modulo solving a Linear Program. Thus, for rational input these two problems are polynomial time equivalent and a polynomial output-sensitive algorithm for one can be used to solve the other in output-sensitive polynomial time. For more details about the problems with current Vertex Enumeration methods, we refer the reader to \cite{DBLP:journals/comgeo/WelzlABS97}.
\medskip

Given a polytope $P\subset \mathbb{R}^d$ the orthogonal projection $\pi(P)$ of $P,$ onto a $k$-dimensional subspace spanned by the first $k$ coordinate directions, is obtained by dropping the last $d-k$ coordinates from every point of $P$. More formally, the projection $\pi: \mathbb{R}^d \rightarrow \mathbb{R}^k$ is a map such that $$\pi(P)=\{x\in\mathbb{R}^k|\exists y\in \mathbb{R}^{(d-k)}, (x,y)\in P\}.$$ In general the projection need not be orthogonal and the projection directions can be an arbitrary orthogonal set of vectors not necessarily aligned with the coordinate axes. In such cases one can apply an affine transform to align the projection directions along the coordinate axes, changing the polytope $P$ in the process to another polytope $P'$ and then considering an orthogonal projection of $P'$.

In this paper we consider the problem of computing the projection of a polytope. Apart from the relation of this problem to the vertex enumeration problem - which we will establish in this paper - another motivation for considering the projection problem arises from the fact that one frequently needs to perform this operation is many areas like Control Theory, Constraint Logic Programming Languages, Constraint Query Languages etc (\cite{DBLP:conf/ppcp/Imbert93}). And although many hardness proofs in this paper are rather simple, they don't appear anywhere in the published literature. Furthermore, many papers seem to indicate that at least in some areas like control theory (See, for example, \cite{RakGri:2004:IFA_2139}) the quest for a polynomial algorithm (even for the versions that we prove to be NP-hard) is still on.

The projection problem has many variants depending on the input representation of the polytope $P$ and the desired output representation of $\pi(P)$. In this paper, we prove that for arbitrary projection directions, most versions of this problem are either equivalent to VE or are NP-hard. Variants that are neither of these admit trivial polynomial algorithms. For example computing the vertices of $\pi(P)$ when $P$ is a $\mathcal{V}$-polytope can be done simply by projecting each vertex and removing redundancies using linear programming. On the other hand, computing the facets or the vertices of $\pi(P)$ when $P$ is an $\mathcal{H}$-polytope is NP-hard (Subsection \ref{arbitrary}). We also review the complexity of the projection problem if the projection directions are in some sense non-degenerate and prove that in many cases one can enumerate the desired form of the output in polynomial time. We would like to note that even though the algorithm for computing non-degenerate projections was conceived independently by the author, it appears to be almost similar to the one presented in \cite{RakGri:2004:IFA_2139}. We nevertheless include it for completeness of discussion.

In order to be able to talk about the equivalence of Vertex Enumeration and projection, we will define a complexity class based on Vertex Enumeration. Keeping in line with other notions of completeness, we call an enumeration problem $\Phi,$ VE-complete if any output-sensitive polynomial algorithm for VE can be used to solve $\Phi$ in output-sensitive polynomial time and vice-versa. Similarly, we call a problem VE-easy if it can be solved in output-sensitive polynomial time using an oracle for VE, and we call a problem VE-hard if an oracle for this problem can be used to solve VE in output-sensitive polynomial time.

The results in this paper are summarized in Table \ref{t1} and Table \ref{t2}. Table \ref{t1} summarizes the complexity of computing projection along arbitrary directions while Table \ref{t2} summarizes the complexity of computing projection along projection directions that satisfy a non-degeneracy criteria.

\begin{table}[htb]

$$\begin{tabular}{|l||l|l|l|}
\hline
Input$\backslash$ Output & $\mathcal{V}$ &$\mathcal{H}$&$\mathcal{HV}$\\
\hline\hline
$\mathcal{V}$& poly & VE-complete &VE-complete \\
\hline
$\mathcal{H}$& NP-hard & NP-hard & VE-complete\\
\hline
$\mathcal{HV}$& poly & VE-complete & VE-complete\\
\hline
\end{tabular}$$

\caption{Complexity of computing projection of a polytope onto an arbitrary subspace
.}
\label{t1}

\end{table}

\begin{table}[htb]

$$\begin{tabular}{|l||l|l|l|}
\hline
Input$\backslash$ Output & $\mathcal{V}$ &$\mathcal{H}$&$\mathcal{HV}$\\
\hline\hline
$\mathcal{V}$& poly & VE-complete & VE-complete \\
\hline
$\mathcal{H}$& VE-hard & poly & VE-complete \\
\hline
$\mathcal{HV}$& poly & poly & poly\\
\hline
\end{tabular}$$
\caption{Complexity of computing projection of a polytope onto a non-degenerate subspace
.}
\label{t2}
\end{table}

Our results about the hardness, and the equivalence of vertex enumeration and computing projection (in most forms) imply that in all forms where the projection can not be computed by a trivial algorithm, finding an output-sensitive polynomial algorithm will be a challenging task. Equivalently, an output-sensitive polynomial algorithm for vertex enumeration will have significant impact in many fields outside algorithmic polytope theory, like Control Theory, Constraint Logic Programming Languages, Constraint Query Languages etc, where one frequently needs to solve the projection problem (\cite{DBLP:conf/ppcp/Imbert93}).

Since for bounded polytopes containing the origin in interior, projection is an operation dual to intersection with a suitably chosen linear subspace, all our results can be dualized by interchanging vertices and facets and replacing projection with intersection.

\medskip

The rest of the paper is organized as follows. In the next section we briefly review some related work about computing projection of a polytope. Our result section is divided into two parts. In Subsection \ref{arbitrary} we present the results about the complexity of computing the projection of a polytope along arbitrary directions and in Subsection \ref{random} we formally state the notion of non-degeneracy of projection directions and describe the complexity results for the case when the projection directions are non-degenerate with respect to the input polytope.


\section{Related Work}

Perhaps the best known algorithm for computing the facets of the projection of an $\mathcal{H}$-polytope is the Fourier-Motzkin elimination discovered by Fourier in 1824 and then rediscovered by Motzkin in 1936. This method is analogous to the method of Gaussian elimination for equations and works by eliminating one variable at a time. Since eliminating one variable from a system of $m$ inequalities can result in $\left \lfloor {\frac{m^2}{4}}\right \rfloor$ facets, the algorithm can have a terrible running time in bad cases where the intermediate polytopes have very large (exponential) number of facets but the final output has only a small number of facets.

Many improvements have been made over the original algorithm (See \cite{DBLP:conf/ppcp/Imbert93} for a survey) but there is no algorithm that has an output-sensitive polynomial running time. The natural question then is whether one can find a shortcut around the intermediate projection steps in the Fourier-Motzkin elimination and obtain an output-sensitive polynomial algorithm. As we will see in Section \ref{results}, the answer is no unless $P = NP$. Thus the lack of any output-sensitive algorithm, for computing the facets of the projection of an $\mathcal{H}$-polytope, is somewhat natural because the problem turns out be NP-hard.

Farka's Lemma provides a way of generating a cone whose extreme rays correspond to the facets of the projection of an $\mathcal{H}$-polytope (\cite{projection_balas}), but unfortunately it does not yield a bijective mapping and many extreme rays of the resulting cone may correspond to redundant inequalities in the projection. Balas \cite{projection_balas} found a way to get rid of these redundancies and provided a way to construct, given an $\mathcal{H}$-polytope $P$, another polyhedral cone $W$ in polynomial time whose projection $W'$ yields a one-to-one correspondence between the extreme rays of $W'$ and the facets of the projection of $P$. Also, $W$ has polynomially many facets compared to $P$.

As noted earlier Jones, Kerrigan and Maciejowski \cite{RakGri:2004:IFA_2139} describe an algorithm for computing the facets of the projection of an $\mathcal{H}$-polytope for non-degenerate projection directions, that is similar to the one presented in this paper. Also, in \cite{AmeZie96} Amenta and Ziegler gave a polynomial algorithm for the case when the input polytope is simple and the projection directions are non-degenerate.

\section{Results}\label{results}
We will denote the projection of a $d$-dimensional polytope $P$ onto a given $k$-dimensional subspace as $\pi_k(P)$. We will mostly omit the subscript and simply refer to the projection as $\pi(P)$ and the projection subspace will depend on the context.

Formally, we are interested in the following problem: Given a polytope $P \in \mathbb{R}^d$ in $\mathcal{H}$-, $\mathcal{V}$- or $\mathcal{HV}$-representation and a set of $k$ orthogonal projection directions defining the projection space, we want to compute the non-redundant $\mathcal{H}$-, $\mathcal{V}$- or $\mathcal{HV}$-representation of $\pi(P)$.

\subsection{Projection onto arbitrary subspaces}\label{arbitrary}
Depending on the input and output form, we have nine variants of the problem. It is obvious that if the vertices are part of the input and one wants to compute the vertices of the projection, then each vertex can be projected trivially and the vertices that become redundant in the projection can be identified by solving one Linear Program per vertex. Hence, we have the following:

\begin{lemma}\label{l1}
 Given a polytope $P \subset \mathbb{R}^d$ in $\mathcal{V}$- or $\mathcal{HV}$-representation and a set of  arbitrary projection directions, non-redundant $\mathcal{V}$-representation of $\pi(P)$ can be computed in polynomial time.
\end{lemma}

Also, it is easy to see that every polytope can be represented as the projection of a suitable simplex. Assuming that the vertices are numbered $0$ through $m-1$, simply append $e_i$ to the $i$-th vertex, where $e_0$ is the zero vector and $e_i$ is the $i$-th unit vector in $\mathbb{R}^{m-1}$. Thus, given a polytope $P$ by its vertices one can compute in polynomial time the vertices of this simplex $\Delta$ and the projection directions such that $P$ is the projection of $\Delta$. Since it is trivial to compute the facets of a simplex given its vertices, Vertex Enumeration can be solved in output-sensitive polynomial time using any algorithm that computes the $\mathcal{H}$- or $\mathcal{HV}$-representation of projection from $\mathcal{V}$- or $\mathcal{HV}$-representation of a polytope.

Clearly, one can also use any polynomial algorithm for Vertex Enumeration to compute the $\mathcal{H}$- or $\mathcal{HV}$-representation of the projection of any polytope given in $\mathcal{V}$- or $\mathcal{HV}$-representation in polynomial time. Hence, we have the following lemma:

\begin{lemma}\label{l2}
Computing the $\mathcal{H}$- or $\mathcal{HV}$-representation of the projection of a polytope given in $\mathcal{V}$- or $\mathcal{HV}$-representation is VE-complete.
\end{lemma}

In what follows, we assume the input polytope $P$ is of the form
$\{(x,y)|Ax+By\leq 1\}$ and we want to compute the projection
$\pi(P)=\{x\in \mathbb{R}^k|(x,y)\in P \text{ for some } y\},$ where
$A\in\mathbb{Q}^{m\times k}, B\in\mathbb{Q}^{m\times (d-k)}, x\in
\mathbb{R}^k, y\in\mathbb{R}^(d-k)$.
We also assume $P$ to be full-dimensional and to contain the origin in
its interior. For rational polytopes this assumption is justified because
one can always find a point in the interior of the polytope via Linear
Programming and move the origin to this point.

We are now left with the three cases where the input polytope is given by $\mathcal{H}$-representation. As we will see now, computing either the facets or the vertices of the projection in this case is hard while computing both facets and vertices of the projection is equivalent to Vertex Enumeration. Consider the following decision version of the problem:\\
\textbf{Input:} Polytopes $P=\{(x,y)|Ax+By\leq 1\}$ and $Q=\{x|A{'}x\leq 1\}$\\
\textbf{Output:} YES if $Q\neq \pi(P)$, NO otherwise.
\medskip

We will now prove that this decision problem is NP-complete thus proving the NP-hardness of the enumeration problem. 

\begin{theorem}\label{pHH-hard}
 Given a polytope $P=\{(x,y)|Ax+By\leq 1\}$ and $Q=\{x|A{'}x\leq 1\}$ it is
NP-complete to decide if $Q\neq \pi(P)$.
\end{theorem}
\begin{proof}
 It is easy to see that deciding whether a given set of hyperplanes completely define the projection of a given higher dimensional polytope, is in NP, since if $Q\neq \pi(P)$ then there is a point that is either only in $Q$ and not in $\pi(P)$ or vice versa. For a given point $x$ checking whether $x\in Q$ is trivial when $Q$ is defined by its facets, and checking whether $x\in\pi(P)$ amounts to checking the feasibility of an LP. So it suffices to show that it is NP-hard as well.

 It is known (\cite{polytope_containment}) that given an $\mathcal{H}$-polytope $P_1=\{x|A^{'}x\leq 1\}$ and a $\mathcal{V}$-polytope $P_2=CH(V)$, it is NP-complete to decide whether $P_1\nsubseteq P_2$. Clearly, $P_1\subseteq P_2$ if and only if $P_1\cap P_2=P_1$. Now, $P_1\cap P_2$ has the following $\mathcal{H}$-representation
\begin{eqnarray*}
A^{'}x&\leq& 1\\
x-\sum_{v\in V}{\lambda_v\cdot v}&=&0\\
\sum_{v\in V}{\lambda_v}&=&1\\
\lambda_v&\geq& 0, \forall v\in V
\end{eqnarray*}

The variables $\lambda$ ensure that we consider only those points in $P$ that can be represented as a convex combination of vertices of $Q$. One can further, in polynomial time, get a full dimensional representation of $P_1\cap P_2$ by eliminating the $d-1$ equations. The resulting polytope is a full-dimensional polytope in $\mathbb{R}^{|V|-1}$ of the form $\{(x,\lambda)|Ax+B\lambda\leq 1\}$.

Since we are interested in only the (vector) variable $x$, the projection of this polytope along the axes corresponding to the variables $\lambda_v$ gives us the facets of $P_1\cap P_2$ in the subspace of variables $x$. It follows that $P_1\cap P_2=P_1$ if and only if the projection of $\{(x,y)|Ax+By\leq 1\}$ onto the subspace of variables $x$ has the $\mathcal{H}$-representation same as that of $P_1$ \textit{i.e.} $A^{'} x\leq 1$.

Thus for arbitrary polytopes $P=\{(x,y)|Ax+By\leq 1\}$ and $Q=\{x|A{'}x\leq 1\},$ an algorithm for deciding whether $Q\neq \pi(P)$ can be used to decide whether an $\mathcal{H}$-polytope is contained in a $\mathcal{V}$-polytope or not, which is a NP-complete problem.
\end{proof}

Balas(\cite{projection_balas}) has shown that for a given $\mathcal{H}$-polytope ($P$) and a set of projection directions, one can compute the facets of another pointed polyhedral cone $W$ and another set of projection directions such that the facets of $\pi(P)$ are in one-to-one correspondence with the extreme rays $\pi(W)$. Note that the projections of $P$ and $W$ are defined in different spaces and should not be confused as the same projection map despite the abuse of notation here. The number of facets of $W$ is polynomial in the number of facets of $P$. It is not difficult to modify the construction in \cite{projection_balas} so that $W$ is bounded \textit{i.e.} a polytope and the vertices in the projection of $W$, except the origin, are in one-to-one correspondence with the facets of the projection of $P$. For completeness we state the result of Balas and describe the modification here.

\begin{lemma}[\textbf{Balas}]\label{l3}
 Given an $\mathcal{H}$ polytope $P$ and a set of projection directions, there exists a polyhedral cone $W$ and another set of projection directions such that the facets of $\pi(P)$, are in one-to-one correspondence with the extreme rays of $\pi(W)$. Furthermore, $W$ has polynomially many facets compared to $P$ and the facets of $W$ can be computed in polynomial time.
\end{lemma}

We will use the notion of polar duality to prove that the cone $W$ obtained from the construction of Balas can be turned into a bounded polytope. For a polyhedral cone $W$ in $\mathbb{R}^n$ with facet inequalities $Ax\leq 0$ and extreme rays the row vectors of $V$, where $A$ and $V$ are matrices with each row a vector in $\mathbb{R}^n,$ the polar dual $W^{*}$ has the roles of the extreme rays and facets reversed. In particular, the facet inequalities of $W^{*}$ are $Vx\leq 0$ and the extreme rays are the row vectors of $A$. We again refer the reader to \cite{grunbaum, ziegler_book} for more details of the properties of polar duality.

\begin{lemma}\label{l4}
 Given a pointed polyhedral cone $W\in \mathbb{R}^n$ and a set of projection directions $\Gamma$ one can construct, in polynomial time, a polytope $P\in \mathbb{R}^{n}$ such that the extreme rays of $\pi(W)$ are in one-to-one correspondence with the vertices of $\pi(P)$, except one vertex corresponding to the apex of $\pi(W)$. Furthermore, both the projections are onto the same subspace.
\end{lemma}
\begin{proof}
Clearly, none of the projection directions lie in the interior of $W$, otherwise the projection spans the whole subspace. Let $W^{*}$ be the polar dual of $W$. For any vector $\alpha$ in the interior of $W^{*}$ the hyperplane $\alpha\cdot x=0 $ touches $W$ only at the origin and hence $W\cap \{x|\alpha\cdot x\leq 1\}$ is a bounded polytope. It is actually a pyramid with origin as the apex.

Now consider the projection $\pi(W)$ which is a pointed cone with origin as apex. This cone is a full-dimensional cone in the subspace containing it and we can consider its polar dual in that subspace. Let $\pi^{*}(W)$ be the polar dual of $\pi(W)$. For any vector $\alpha^{'}$ in the interior of $\pi^{*}(W)$, $\pi(W)\cap \{x|\alpha^{'}\cdot x\leq 1\}$ is a bounded polytope. Moreover, for such an $\alpha^{'}$, $\gamma_i\cdot \alpha^{'} =0$ for all $\gamma_i \in \Gamma$. Since $\pi(W)$ is a pointed cone such an $\alpha^{'}$ exists.

Since $\pi^{*}(W)$ can be obtained as the intersection of $W^{*}$ with $\bigcap_{\gamma_i \in \Gamma}{\{\gamma_i\cdot x=0\}}$, a vector $\alpha^{'}$ in the interior of $\pi^{*}(W)$ can be computed in polynomial time if one knows either the vertices or facets of $W$. Also, $\alpha^{'}$  lies in the interior of $W^{*}$ as well. Define $Q=W\cap \{\alpha^{'}\cdot x\leq 1\}$. Given the extreme rays (facets respectively) of $W$ and the projection directions $\Gamma$ one can compute the vertices (facets respectively) of $Q$ in polynomial time.

Since $\alpha^{'}$ is orthogonal to each of the projection directions, the vertices and facets of $\pi(P)$ are in one-to-one correspondence with the extreme rays and the facets of $\pi(W)$. Note that one vertex of $\pi(P)$ corresponds to the apex of $\pi(W)$.
\end{proof}

Theorem \ref{pHH-hard} together with Lemma \ref{l3} and Lemma \ref{l4} gives the following:

\begin{theorem}
 Given a polytope $P=\{(x,y)|Ax+By\leq 1\}$ and $Q=CH(V)$ it is NP-complete to decide if $Q\neq \pi(P)$.
\end{theorem}

Now we consider the last variant of the projection problem where we are given an $\mathcal{H}$-polytope and we want to compute the $\mathcal{HV}$-representation of the projection. It turns out that although computing either the vertices or facets of the projection is NP-hard, computing both vertices and facets is VE-complete.

Before we prove this, we would like to remark that the notion of output-sensitiveness can have various meanings. An output-sensitive polynomial algorithm for an enumeration problem (like VE) could enumerate vertices such that a new vertex is reported within incremental polynomial delay \emph{i.e.} each new reporting takes time polynomial in the input and the output produced so far. It is equally conceivable that the algorithm takes total time polynomial in the input and output but there is no guarantee that successive reportings take only incremental polynomial delay. We will assume that if we have an output-sensitive algorithm of the latter kind, then we actually know the complexity of its running time. Under this assumption the two notions are same for VE.

{\small{To see why this is true, consider the following. Given an $\mathcal{H}$-polytope $P$ and a $\mathcal{V}$-polytope $Q$, determining whether $P=Q$ is polynomial time equivalent to VE (See \cite{DBLP:journals/comgeo/WelzlABS97}). Also, solving this problem gives an algorithm for VE that is not only output-sensitive polynomial but also has a polynomial delay guarantee. If we have an enumeration algorithm that has no guarantee of polynomial delay between successive outputs, but for which we know the running time, then we can use this procedure to create a polynomial algorithm for deciding the equivalence of $\mathcal{H}$- and $\mathcal{V}$-polytopes: Simply compute the time needed by the algorithm to enumerate all vertices of $P$ assuming $P=Q$ and run the enumeration algorithm for the time required to output $|vert(Q)|+1$ vertices. If the procedure stops then we can compare the list of vertices of $P$ with that of $Q$ in polynomial time. If, on the other hand, the procedure doesn't finish within the given time then $P$ must have more vertices than $Q$ and hence, $P\neq Q$. This also implies that if we have an algorithm for VE that has output-sensitive polynomial running time, then we can assume that the algorithm produces successive vertices with only a delay polynomial in the size of the input and the number of vertices produced so far.}}

So for proving VE-completeness in the next theorem, when we assume the existence of an output-sensitive polynomial algorithm for VE, we also assume that this algorithm has a guarantee of polynomial delay between successive outputs. Although we will work with the Convex Hull problem which is the dual version of VE, with a slight abuse of language we will refer to this dual problem as VE as well.

\begin{theorem}
 Given a polytope $P=\{(x,y)|Ax+By\leq 1\}$ it is VE-complete to compute the facets and vertices of $\pi(P)$.
\end{theorem}
\begin{proof}
 Since every polytope $P \in \mathbb{R}^n$ given by $m$ vertices can be converted to a $(m-1)$-dimensional simplex $\Delta$ such that $P$ is a projection of $\Delta$ it is clear that computing $\mathcal{HV}$-representation of the projection of an $\mathcal{H}$-polytope is VE-hard.

To prove that this problem is also VE-easy, we give an algorithm that uses a routine for VE to enumerate the facets and vertices of $\pi(P)$. The algorithm proceeds as follows: At any point we have a list of vertices $V$ of $\pi(P)$ and we want to verify that $V$ indeed contains all vertices of $\pi(P)$. If the list is not complete, we want to find another vertex of $\pi(P)$ that is not already in $V$. To do this, we start enumerating facets of $CH(V)$ and we verify that each generated facet is indeed a facet of $\pi(P)$. Note that even though $CH(V)$ can have many more facets than $\pi(P)$ (in fact it can have exponentially many facets compared to $\pi(P)$ \cite{DBLP:journals/comgeo/WelzlABS97}), this is not a problem since we stop at the first facet of $CH(V)$ that is not a facet of $\pi(P)$. Checking whether a given facet of $CH(V)$ is a facet of $\pi(P)$ or not is easy because of the following:

Suppose $h=\{x|a\cdot x=1\}$ be a hyperplane in the projection space. We say that $h$ intersects $P$ properly if the intersection $P\cap \{(a,\overbrace{0,\cdots,0}^{k~ \text{times}})\cdot(x,y)=1\}$ has some point in the interior of $P$. We will call such an intersection a \textit{proper} intersection.

We claim that the defining hyperplane of every facet $f$ of $CH(V)$, that is not a facet of $\pi(P)$, intersects $P$ properly. To see this, pick a point $x_1$ in the relative interior of $f$. Such a point exists because $CH(V)\subset \pi(P)$ if some facet $f$ of $CH(V)$ is not a facet of $\pi(P)$. This point also lies in the relative interior of $\pi(P)$. Also, there is a point $(x_1,y_1)$ that lies in the interior of $P$ that projects to $x_1$. Clearly the hyperplane $\{(a,\overbrace{0,\cdots,0}^{k~ \text{times}})\cdot(x,y)=1\}$ contains $(x_1,y_1)$ and hence the hyperplane defining $f$ intersects $P$ properly.

It follows that, if $V$ does not contain all vertices of $\pi(P)$ then there exists a facet $f=\{x|a\cdot x=1\}$ of $CH(V)$ intersecting $P$ properly. So if the enumeration procedure for facets of $CH(V)$ stops and none of the facets intersect $P$ properly then $V$ contains all the vertices of $\pi(P),$ and we have all the vertices and facets of $\pi(P)$. If some intermediate facet $\{a\cdot x=1\}$ of $CH(V)$ does intersect $P$ properly then maximizing the objective function $(a,\overbrace{0,\cdots,0}^{k~ \text{times}})\cdot(x,y)$ over $P$ produces a vertex of $P$ that also gives a vertex $v$ of $\pi(P)$ upon projection. Moreover this vertex is not in the list $V$. This gives an output-sensitive polynomial algorithm for enumerating all facets and vertices of $\pi(P)$. Hence, computing all vertices and facets of the projection $\pi(P)$ of an $\mathcal{H}$-polytope $P$ is VE-easy as well.
\end{proof}


%
%


\subsection{Projection onto non-degenerate subspaces}\label{random}

To make the notion of non-degenerate projection precise, note that if $P$ is the input polytope then every face of projection $\pi(P)$ is the shadow of some proper face of $P$. Call the maximal dimensional face $f'$ of $P$ a \textit{pre-image} of the face $f$ of $\pi(P)$ if $f$ is obtained by projecting all vertices defining $f'$ and taking their convex hull. In general, the dimensions of $f$ and $f'$ are not the same. This can happen if some projection directions lie in the affine hull of $f'$. We call a set of projection directions \textit{non-degenerate} with respect to $P$ if no directions lie in the affine hull of any face of $P$.

\medskip
\noindent \textbf{Fact 1:} For non-degenerate projection directions and a face $f$ of $\pi(P)$, if $f'$ is the pre-image of $f$ then $dim(f)=dim(f')$.
\medskip

This is easy to see because for any face $f'$ of $P$ that appears in $\pi(P)$ projection reduces the dimension if and only if the some projection directions lie in the affine hull of $f'.$ This is not possible for non-degenerate projection directions.


Now, given a polytope $P$ in $\mathcal{H}$-representation and a set of non-degenerate projection directions $\Gamma$ we want to compute the facets of the projection $\pi(P)$. Again, we assume that the facets of $P$ are presented as inequalities of the form $Ax+By\leq 1$, where $A$ is an $m \times k$ matrix, $B$ is an $m \times (d-k)$ matrix, and the projection has dimension $k$. Since, we will need to solve Linear Programs we also assume that the polytope is rational \textit{i.e.} the entries in $A$ and $B$ are rational numbers. We will assume that the projection directions are aligned along a subset of co-ordinate axes. If not, we can apply a suitable affine transform to $P$ depending on the orthogonal projection directions.


Our algorithm for enumerating the facets of $\pi(P)$ proceeds as follows: Given a partial list of facets of $\pi(P)$, for each facet $f$ we identify its pre-image $f'$ in $P$. For each of these faces of $P$ we identify their $(d-2)$-dimensional faces and among all such $(d-2)$-faces of $f'$ some give rise to ridges in $\pi(P)$. We identify which faces form the pre-image of some ridge of $\pi(P)$ and from the corresponding ridge, we identify the two facets defining this ridge, thus finding a new facet of $\pi(P)$ if the current list of facets is not complete.

\begin{lemma}{\label{l5}}
 Given an $\mathcal{H}$-polytope $P$ and a facet $f$ of its projection $\pi(P)$, one can find the facets of $P$ defining the pre-image of $f$ in polynomial time.
\end{lemma}
\begin{proof}
 Let $\{x\in \mathbb{R}^d|a\cdot x \leq 1, a\in \mathbb{Q}^d\}$ be the halfspace defining the facet $f$ of $\pi(P)$. Clearly, the hyperplane $h$ in $\mathbb{R}^{d+k}$ with normal $a'=(a,\overbrace{0,\cdots,0}^{k \text{ times}})$ defines the supporting hyperplane $\{x\in \mathbb{R}^{d+k}|a'\cdot x = 1\}$. Also, $P\cap h$ is a face of $P$ and is exactly the pre-image of $f$. A facet $F$ of $P$ contains this face iff $P\cap h \cap F$ has the same dimensions as $P\cap h$. Thus, one can find all the facets of $P$ containing the pre-image of $f$ in time polynomial in the size of $P$.
\end{proof}

The next lemma follows immediately from the non-degeneracy of the projection directions, so we mention it without the proof (See Fact 1).

\begin{lemma}{\label{l6}}
 Given $P$ and a facet $f$ of its projection $\pi(P)$, if $g$ is another facet of $\pi(P)$ sharing a ridge with $f$ then the pre-images $f'$ and $g'$ share a face in $P$. Furthermore, $$dim(f'\cap g')=dim(f\cap g)=dim(f')-1=dim(g')-1=d-2$$
\end{lemma}

Since the facets of $P$ are known, we can identify all $(d-2)$-faces of $f'$. The number of these faces is at most $m$ for each pre-image $f'$ and since $f'$ is itself a polytope of dimension $d-1$, we can compute the non-redundant inequalities defining the facets ($d-2$-dimensional faces) of $f'$. At this point, what remains is to identify these ridges and the facets defining these ridges. The following two lemmas achieve this.

\begin{lemma}{\label{l7}}
 Let $P=\{(x,y)|Ax+By\leq 1\}$ be a polytope in $\mathbb{R}^d$, where $A\in \mathbb{Q}^{m\times k}, B\in \mathbb{Q}^{m\times (d-k)}$ $x\in \mathbb{R}^{k}, y\in \mathbb{R}^{(d-k)}$. Also, let $f$ be a $(d-2)$-face of $P$ defined as $f=\{(x,y)|A^{'}\cdot (x,y)=1, (x,y)\in P\}$. Then, $f$ defines a ridge in the projection $\pi(P)$ if and only if
\begin{itemize}
 \item there exists $\alpha \in \mathbb{R}^d$ such that $(\alpha,\overbrace{0,\cdots,0}^{k\text{ times}}) \in CH(A^{'})$, where each row of $A^{'}$ is treated as a point in $\mathbb{R}^{d+k}$. And,
 \item The feasible region of all such $\alpha$ is a line segment.
\end{itemize}
\end{lemma}

 It is not difficult to see that this lemma is just a rephrasing of the basic properties of supporting hyperplanes of a polytope. In other words, any hyperplane whose normal is a convex combinations of the normals of facets defining the face $f$, is a supporting hyperplane of $P$ and vice-versa. Furthermore, if the normal lies in the subspace where the projection $\pi(P)$ lives, then it is also a supporting hyperplane of $\pi(P)$. Also, the normals of all hyperplanes that support a polytope at some ridge, when treated as points, form a $1$-dimensional polytope \textit{i.e.} a line segment. This formulation allows us to check in polynomial time whether a $(d-2)$-face of $P$ forms a pre-image of some ridge of $\pi(P)$.

\begin{lemma}{\label{l8}}
 The end points of the feasible region of $\alpha$ in lemma \ref{l7} are the normals of the facets of $\pi(P)$ defining the ridge corresponding to face $f$.
\end{lemma}
 As noted before, the normals of the hyperplanes supporting a polytope at a ridge $r$ form a line segment when viewed as points. The end points of the segment represent the normals of the two facets defining the ridge $r$. This lemma ensures that given a pre-image of some ridge of $\pi(P)$, one can compute the normals of the two facets of $\pi(P)$ defining the ridge $r$ by solving a polynomial number of linear programs each of size polynomial in the size of input.

Putting everything together we get the following theorem:

\begin{theorem}\label{t4}
 Given a polytope $P$ defined by facets, and a set of non-degenerate orthogonal projection directions $\Gamma$ one can enumerate all facets of $\pi(P)$ in output-sensitive polynomial time.
\end{theorem}

Since randomly picked projection directions are non-degenerate with probability $1$, one can also enumerate the facets of the projection of an $\mathcal{H}$-polytope for such directions. Note, that this also gives an output-sensitive polynomial algorithm for the case when the input is an $\mathcal{HV}$-polytope irrespective of the output form. Also, if the vertices of $P$ are given then some tests like those in Lemma \ref{l7} and \ref{l8} become easier. We leave the proof of this to the reader since they do not affect our main argument about the existence of an output-sensitive polynomial algorithm.

\begin{corollary}
 Given an $\mathcal{HV}$-polytope $P$ and a set of projection directions $\Gamma$ that are non-degenerate with respect to $P$ there is an algorithm that can enumerate the vertices and/or facets of $\pi(P)$ in output-sensitive polynomial time.
\end{corollary}

It is easy to see that computing the vertices of the projection of an $\mathcal{H}$-polytope along non-degenerate directions has VE as a special case - simply pick the set of projection directions to be the empty set. Hence, enumerating vertices of the projection of an $\mathcal{H}$-polytope along non-degenerate projection directions remains VE-hard even though it is not clear if it remains NP-hard. Similarly one can argue that the complexity status of computing the projection of a $\mathcal{V}$-polytope along non-degenerate projection directions remains the same as that of computing the projection along arbitrary directions. 
%

\section{Concluding Remarks}
Computing the projection of a polytope is a fundamental task that arises frequently in many different areas of computer science like Control Theory, Constraint Logic Programming Languages, Constraint Query Languages among others \cite{DBLP:conf/ppcp/Imbert93}. The results in this paper emphasize the importance of obtaining an efficient method for vertex enumeration for computing projections. We also defined the notions of VE-hardness, completeness etc in this paper. Given the long history of research on the vertex enumeration and lack of any insight on its complexity, we believe that it is natural to try to relate the complexity of new problems with that of vertex enumeration. We hope that the new complexity classes defined in this paper will be useful for future research and new problems will be shown to be in these classes. 
\bibliographystyle{abbrv}
\bibliography{projection}
\end{document}